\documentclass[10pt,letterpaper]{article}
\usepackage[top=0.85in,left=2.75in,footskip=0.75in,marginparwidth=2in]{geometry}

\usepackage[utf8]{inputenc}

\usepackage[style=numeric]{biblatex}
\usepackage{nameref,hyperref}

\usepackage[right]{lineno}

\usepackage{microtype}
\DisableLigatures[f]{encoding = *, family = * }

\raggedright
\usepackage{changepage}

\usepackage[aboveskip=1pt,labelfont=bf,labelsep=period,singlelinecheck=off]{caption}

\makeatletter
\renewcommand{\@biblabel}[1]{\quad#1.}
\makeatother

\usepackage{lastpage,fancyhdr,graphicx}
\usepackage{epstopdf}
\fancyheadoffset[L]{2.25in}
\fancyfootoffset[L]{2.25in}

\usepackage{color}

\definecolor{Gray}{gray}{.25}

\usepackage{graphicx}
\usepackage{subcaption}

\usepackage{amsmath,amssymb,amsfonts,amsthm}

\usepackage{hyperref}

\usepackage{sidecap}

\usepackage{wrapfig}
\usepackage[pscoord]{eso-pic}
\usepackage[fulladjust]{marginnote}
\reversemarginpar

\begin{document}
\vspace*{0.35in}

\begin{flushleft}
{\Large
\textbf\newline{A Closed-Form Solution to the 2-Sample Problem for Quantifying Changes in Gene Expression using Bayes Factors}
}
\newline
\\
Franziska Hoerbst\textsuperscript{*},
Gurpinder Singh Sidhu\textsuperscript{},
Melissa Tomkins\textsuperscript{},
Richard J. Morris\textsuperscript{*}
\\
\bigskip
Computational and Systems Biology, John Innes Centre, Norwich Research Park, UK
\\
\bigskip
* Franziska.Hoerbst@jic.ac.uk, Richard.Morris@jic.ac.uk

\end{flushleft}

\section*{Abstract}
Sequencing technologies have revolutionised the field of molecular biology. We now have the ability to routinely capture the complete RNA profile in tissue samples. This wealth of data allows for comparative analyses of RNA levels at different times, shedding light on the dynamics of developmental processes, and under different environmental responses, providing insights into gene expression regulation and stress responses. However, given the inherent variability of the data stemming from biological and technological sources, quantifying changes in gene expression proves to be a statistical challenge. Here, we present a closed-form Bayesian solution to this problem. Our approach is tailored to the differential gene expression analysis of processed RNA-Seq data. The framework unifies and streamlines an otherwise complex analysis, typically involving parameter estimations and multiple statistical tests, into a concise mathematical equation for the calculation of Bayes factors. Using conjugate priors we can solve the equations analytically. For each gene, we calculate a Bayes factor, which can be used for ranking genes according to the statistical evidence for the gene's expression change given RNA-Seq data. The presented closed-form solution is derived under minimal assumptions and may be applied to a variety of other 2-sample problems.




\section{Introduction}
\begin{figure*}[h!]
    \centering
    \begin{subfigure}[]{\textwidth}
        \centering
        \includegraphics[width=\textwidth]{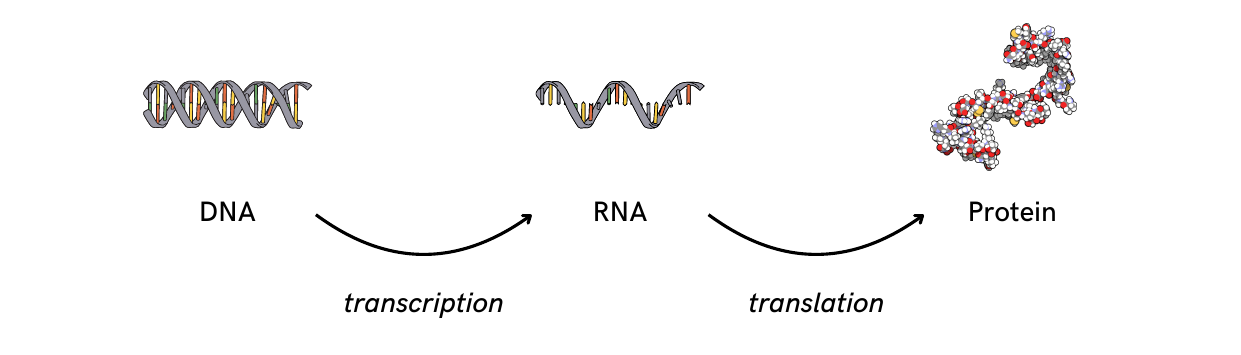}
    \end{subfigure}
    \vspace{5mm}
    \caption[]{Transcription and translation are highly dynamic, tightly regulated processes. Quantifying changes of molecular populations over time can provide important insights into regulatory processes in the cell.  
    }
    \label{fig:CENTRALDOGMA}
\end{figure*}

Every living cell of any organism on this planet follows the so-called central dogma of Molecular Biology -- a genetic information processing pipeline and molecular production pathway, Figure \ref{fig:CENTRALDOGMA}. The processes by which genetic information, stored in DNA, is transcribed into RNA and translated into protein are highly dynamic. Transcription and translation are intricately regulated; quantifying these dynamic processes can shed light on the underlying regulation. Significant effort is therefore being invested in measuring changes in RNA and protein levels throughout development and under different experimental perturbations. \\

\begin{figure*}[h!]
    \centering
    \begin{subfigure}[]{\textwidth}
        \centering
        \includegraphics[width=\textwidth]{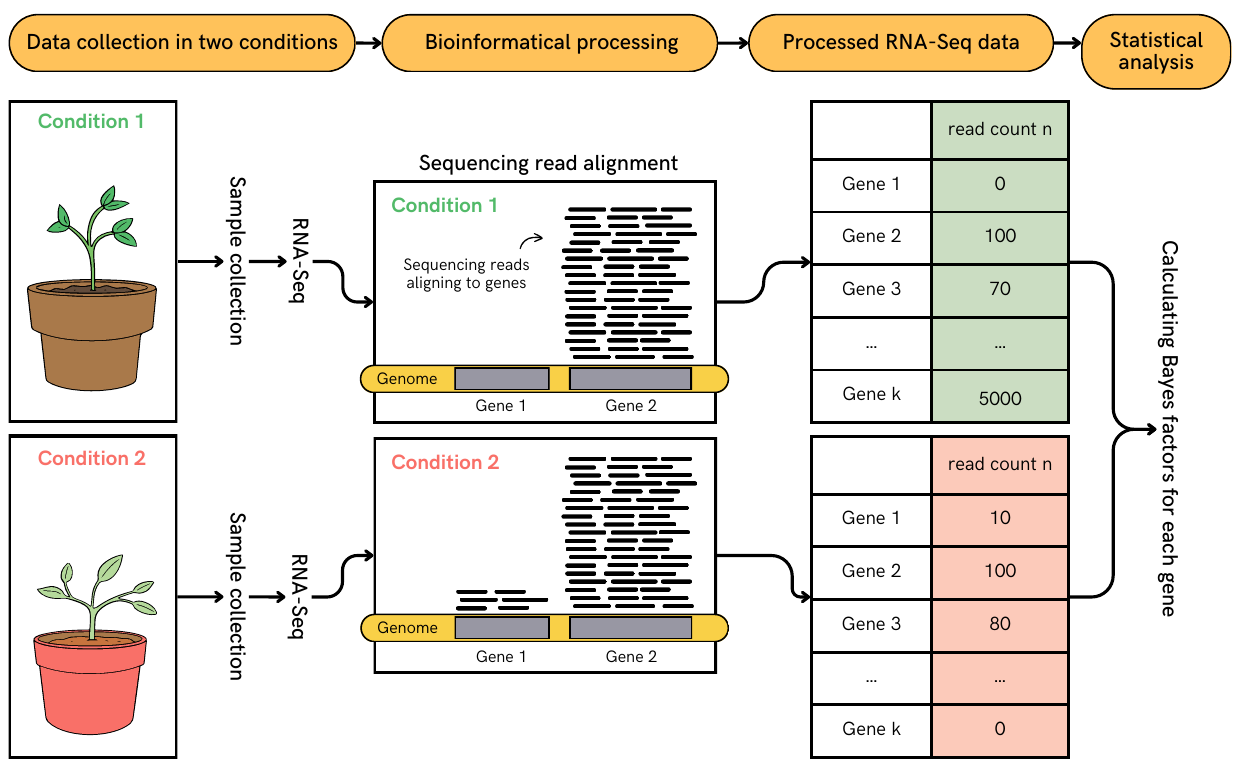}
    \end{subfigure}
    \vspace{5mm}
    \caption[]{An overview of RNA-Sequencing experiments to identify differentially expressed genes. Tissue samples are collected from two conditions of interest. The RNA in these samples is extracted separately, and sequenced. The raw sequencing data output is processed by bioinformatics pipelines. One important step is the alignment of sequencing reads to the genome assembly. The processed RNA-Seq data consists of sequencing read counts for each gene. The statistical analysis follows, which in this case is to find out which genes have changed their expression between the conditions. Our solution for the 2-sample problem is to calculate a Bayes factor for differential gene expression for each gene to rank genes according to the evidence in the data for changes in the amounts of RNA present in the samples.}
    \label{fig:1}
\end{figure*}

 The main high-throughput method for measuring transcript levels in cells is RNA-Sequencing \cite{Wang_Snyder_2009, Behjati_Tarpey_2013, Furlan_Pelizzola_2020}. In this technique, a biological sample is taken (e.g. a leaf of a plant), and by chemical and physical treatments, all RNA of all cells in the sample is extracted, while all other structures and molecules are washed away. The extracted RNA is sequenced, which results in short strings of RNA (sequencing reads). These reads can be compared (aligned) to the genome sequence of the organism to find which gene was transcribed. For each gene in the genome, we can count the number of reads mapping to it. This number is assumed to be proportional to the amount of RNA originating from this gene in the tissue. Hence, by sequencing all RNAs found in a biological sample, we can determine the amount of RNA in a tissue. If we carry out two (or more) RNA-Sequencing experiments under different conditions, we can compare the amounts of RNA in a sample and may learn about environmental conditions affecting the process of transcription in a tissue. In differential gene expression analysis, the goal is to identify genes with a `significant' change in their expression between the two measured conditions (`differentially expressed genes' or DEG). In Figure \ref{fig:1} we present an overview of this experimental and computational procedure. \\

Currently popular statistical tools \cite{Su_SEQC/MAQC-IIIConsortium_2014, Anders_Robinson_2013, Zhou_Robinson_2014, Conesa_Mortazavi_2016, VandenBerge_Robinson_2019a, Corchete_Burguillo_2020, Koch_Winter_2018,  Costa-Silva_Lopes_2017, Costa-Silva_Lopes_2023, McDermaid_Ma_2019, Stark_Hadfield_2019,Schurch_Barton_2016, Chen_Marquez-Lago_2023, Rapaport_Betel_2013, Love_Anders_2014} rely on two criteria for determining a DEG: (1) p-values below a set threshold (for normalised read counts between the two conditions) and (2) absolute $\log_2$ fold change above a certain threshold. Typical cutoffs for DEGs are p-value $<$ 0.05, and $\lvert \ \log_2$ (fold change) $\rvert >1$. 
While setting a fold change cutoff decreases the number of false positive hits, potentially interesting genes with noticeable changes that have not at least halved or doubled their expression, are ignored. However, there is no reason why the impact of a gene that has doubled its RNA in a cell is necessarily higher than another gene that has incurred a smaller change. As an alternative to deciding arbitrary fold change cut-offs, we propose a Bayesian framework tailored to the differential gene expression analysis of processed RNA-Seq data. In contrast to current software that often involves various pre-processing stages, and applies complex statistical analyses, involving parameter estimations and multiple statistical tests, we put forward a concise mathematical equation (i.e. we provide a closed-form or analytical solution) to calculate a Bayes factor for each gene, enabling genes to be ranked according to the statistical evidence for change.

\section{Results and Discussion}
\subsection{Differential gene expression analysis 
can be cast into the framework of Bayesian inference and model comparison}

We assign to every gene $i$ an expression probability $q_i$. This probability summarises all events from the transcription of gene $i$ to the mapping of a corresponding read in data processing, 

\begin{equation}
    {\rm gene} \quad \xrightarrow{q_i} \quad {\rm read \ count}.
\end{equation}

Importantly, this expression probability $q_i$ can change over time and in response to external stimuli. For a known $q_i$, we can describe the probability of $n_i$ number of reads mapping to a gene $i$, out of total reads in a sample $N$ with a binomial distribution (a read maps to gene $i$, with probability $q_i$, and it does not, with probability $1-q_i$),

\begin{equation} \label{binomiald}
    P(n_i|N,q_i) 
    = {N\choose{n_i}} \ q_i^{n} (1-q_i)^{N-n_i}.
\end{equation}

This implies that if we know the gene expression probability $q_i$ of a gene and the total number of all reads $N$ in an RNA-Seq experiment, we can compute the probability of any number of RNA-Seq reads $n_i$ mapping to gene $i$. Note that $N = \sum n_i$. \\

In a differential gene expression experiment (or any other two-sample test) we want to answer the question of whether two data sets $D_1$ (consisting of $N_1$ and $n_{i1}$) and $D_2$ (consisting of $N_2$ and $n_{i2}$), for an observation (gene) $i$, \\
\begin{equation}
    q_{i1} \leadsto D_1 \quad {\rm and} \quad q_{i2} \leadsto D_2
\end{equation}
arose from the same probability distribution $q_i$. Therefore, our problem is to decide whether

\begin{equation}
    q_{i1} \quad \overset{?}{=} \quad q_{i2},
\end{equation}

which is equivalent to asking whether the expression probability $q_i$ of a gene changed between two data sets, or not. \\

We can calculate a Bayes factor for each gene, describing how much the RNA-Seq data supports one of two hypotheses. This will be our metric for ranking genes according to the statistical evidence for a change in gene expression. \\ 

We introduce two hypotheses. 
{\bf Hypothesis $\bf H_1$} assumes that the data from both experiments can be explained by one statistical model, 
    \begin{equation}
        q_i \leadsto D_1 \quad {\rm and} \quad q_i \leadsto D_2. 
    \end{equation}
    
Having a common $q_i$ value means that the RNA-Seq data from the first and second experiment, i.e. mRNA levels of gene $i$, are consistent. \\

In {\bf Hypothesis $\bf H_2$} the data are best explained by a different model for each experiment,  
    \begin{equation}
        q_{i1} \leadsto D_1 \quad {\rm and} \quad q_{i2} \leadsto D_2.
    \end{equation}

If there is more statistical evidence for hypothesis 2 than hypothesis 1, the data support a difference in gene expression between the samples. \\

\subsection{Bayes factors for differential gene expression can be computed analytically}

The $q_i$ values are not known,  however, we can infer them from RNA-Seq data using Bayes' theorem. We use $\theta_i$ to denote the continuum of possible values of $q_i$ and use a probability distribution over $\theta_i$, $P(\theta_i)$, to capture our knowledge of $q_i$, with our best estimate of $q_i$ being the expectation value $\langle \theta_i \rangle$.  \\


To compute the posterior distribution,

\begin{equation} \label{PthetanN}
    \addtolength{\jot}{1em}
    \begin{split}
        P(\theta_i|D) & \ = \ \frac{P(D|\theta_i) \times P(\theta_i)}{P(D)},
    \end{split}
\end{equation}
we need to determine the three terms. 


To find a likelihood function $P(D|\theta_i)$ for all possible $\theta_i$, we can use the equation \ref*{binomiald} to model $P(D|\theta_i)$ as a simple binomial process,

\begin{equation}
    P(D|\theta_i) \ \propto \ {N\choose{n_i}} \ \theta_i^{n_i}(1-\theta_i)^{N-n_i}.
\end{equation}

We choose a conjugate prior (the Beta distribution), allowing us to proceed analytically. The prior distribution can thus be written as 

\begin{equation} \label{Betadistribution}
    P(\theta_i|u_1,u_2) \ = \ \frac{1}{B(u_1,u_2)} \ \theta_i^{u_1-1} \ (1-\theta_i)^{u_2-1} \ = \ {\rm Beta}(u_1,u_2),
\end{equation}

with hyper-parameters $u_1, u_2 \in \mathbb{R}_{>0}$. Here, we choose a bias-free, flat prior ($u_1, u_2 = 1$). \\

Finally, the evidence $P(D)$ which can be expressed as 
\begin{equation}
    P(D) \ = \ \int_0^1 P(D|\theta_i) \times P(\theta_i) \ d\theta_i.
\end{equation}


For two datasets, $D_1$ and $D_2$, we can find a posterior distribution $P(\theta_i|D)$ for Hypothesis 1 and Hypothesis 2 separately. For Hypothesis 1, the assumption is $\theta_i \leadsto D_1$ and $\theta_i \leadsto D_2$, resulting in

\begin{equation}
    \addtolength{\jot}{1em}
    \begin{split}
        P(\theta_i|D_1, D_2, H_1) \quad & = \quad \frac{P(D_1, D_2|\theta_i, H_1) \times P(\theta_i|H_1)}{P(D_1, D_2|H_1)}.
    \end{split}
\end{equation}

For Hypothesis 2, where $\theta_{i1} \leadsto D_1$ and $\theta_{i2} \leadsto D_2$

\begin{equation} \label{Ptheta1theta2D1D2H2}
    \addtolength{\jot}{1em}
    \begin{split}
        P(\theta_{i1}, \theta_{i2}|D_1, D_2, H_2) \quad &= \quad \frac{P(D_1, D_2|\theta_{i1}, \theta_{i2}, H_2) \times P(\theta_{i1}, \theta_{i2}|H_2)}{P(D_1, D_2|H_2)}.
    \end{split}
\end{equation}

We define a prior for the model with a single parameter (Hypothesis 1),

\begin{equation}
    P(\theta_i|H_1) \ = \ {\rm Beta}(u_1,u_2) \ = \ \frac{1}{B(u_1,u_2)} \ \theta_i^{u_1-1} \ (1-\theta_i)^{u_2-1}
\end{equation}

and the prior for two parameter model (Hypothesis 2),

\begin{equation}
    P(\theta_{i1}, \theta_{i2}|H_2) \ = \ P(\theta_{i1}|H_2) \times P(\theta_{i2}|H_2) \ = \ {\rm Beta}(u_1,u_2) \times {\rm Beta}(u_1,u_2).
\end{equation}

Note that we have assumed the same priors over $\theta_{i1}$ and $\theta_{i2}$. The likelihood for the data ($D_1, D_2$) given Hypothesis 1 can be written as \\

\begin{equation} \label{PthetaiD1D2H1}
    \addtolength{\jot}{1em}
    \begin{split}
        P(D_1, D_2| \theta_i, H_1) \ = \ & {N_1\choose{n_{i1}}} \theta_i^{n_{i1}} (1-\theta_i)^{N_1 - n_{i1}} \ \times \ {N_2\choose{n_{i2}}} \theta_i^{n_{i2}} (1-\theta_i)^{N_2 - n_{i2}}. 
    \end{split}
\end{equation}

For Hypothesis 2, the likelihood of the data given $H_2$, depends on the two parameters, $\theta_{i1},\theta_{i2}$, respectively,

\begin{equation} \label{Pthetai1thetai2D1D2H2}
    \addtolength{\jot}{1em}
    \begin{split}
        P(D_1, D_2| \theta_{i1}, \theta_{i2}, H_2) \ = \ & {N_1\choose{n_{i1}}} \theta_{i1}^{n_{i1}} (1-\theta_{i1})^{N_1 - n_{i1}} \ \times \ {N_2\choose{n_{i2}}} \theta_{i2}^{n_{i2}} (1-\theta_{i2})^{N_2 - n_{i2}}.
    \end{split}
\end{equation}

Thus, the posterior of Hypothesis 1 simplifies, thanks to conjugate priors, to the following equation,

\begin{equation} \label{PthetaiD1D2H1}
    \addtolength{\jot}{1em}
    \begin{split}
        P(\theta_i|D_1, D_2, H_1) \quad = \quad & \frac{ P(D_1, D_2|\theta_i, H_1) \times P(\theta_i|H_1)}{P(D_1, D_2|H_1)} \quad = \\
        = \quad & {\rm Beta}(u_1+n_{i1}+n_{i2},u_2+N_1+N_2-n_{i1}-n_{i2}).
    \end{split}
\end{equation}

Note how the evidence (denominator) simplifies to a Beta function and how the posterior can be expressed as a Beta distribution, compare Equation \ref{Betadistribution}. \\

Analogously, for Hypothesis 2 we can formulate and simplify the expression for the posterior probability distribution to

\begin{equation} \label{Pthetai1thetai2H2D}
    \addtolength{\jot}{1em}
    \begin{split}
    P(\theta_{i1},\theta_{i2}|D_1, D_2, H_2) \quad = \quad & \frac{P(D_1, D_2|\theta_{i1},\theta_{i2}, H_2) \times P(\theta_{i1}, \theta_{i2}|H_2)}{P(D_1, D_2|H_2)} \quad = \\
    = \quad & {\rm Beta}(u_1+n_{i1},u_2+N_1-n_{i1}) \ \times \\ 
    & {\rm Beta}(u_1+n_{i2},u_2+N_2-n_{i2}).
    \end{split}
\end{equation}

We can compare hypotheses using the posterior odds ratio, $P(H_2|D)/P(H_1|D)$ \cite{Jaynes_Jaynes_2003,Sivia_Skilling_2006}. When both hypotheses, $H_1$ and $H_2$, are equally likely \textit{a priori}, $P(H_1) = P(H_2)$, the posterior odds ratio becomes equal to the marginal likelihood or evidence ratio. This evidence ratio is the Bayes factor. We already derived expressions for the evidences (the Beta functions in denominator of the posterior), allowing us to express the Bayes factor as 

\begin{equation} \label{BF}
\addtolength{\jot}{1em}
\begin{split}
    BF \quad = & \quad \frac{P(D_1, D_2|H_2)}{P(D_1, D_2|H_1)} \quad = \\
    = & \quad \frac{B(u_1+n_{i1},u_2+N_1-n_{i1}) \ \times \ B(u_1+n_{i2},u_2+N_2-n_{i2})}{ B(u_1,u_2) \ \times \ B(u_1+n_{i1}+n_{i2},u_2+N_1+N_2-n_{i1}-n_{i2}),}
\end{split}
\end{equation}

where all except one pre-factor, $B(u_1,u_2)$, cancel. For a flat prior, $B(u_1,u_2) = 1$. \\  

As is common practice in differential gene expression analysis, we proceed to calculate an inferred $\log_2$ fold change ($iFC$). \\

\begin{equation}
    iFC = \log_2 \biggl\{ \ \Bigl( \frac{u_1 + n_2}{u_2 + N_2 - n_2} \Bigl) \ / \ \Bigl( \frac{u_1 + n_1}{u_2 + N_1 - n_1} \Bigl) \  \biggl\}
\end{equation}

Under the assumption that replicates $r_j$ for each condition $j$ are consistent with each other (which can be evaluated using the same framework), the $N_{r_j}$ of all replicates can be summed to $N_j$ for each condition $j$ and all $n_{i,r_j}$ to $n_{i,j}$ to calculate $BF$ and $iFC$ for each gene $i$. \\

\subsection{Genes can be ranked according to the statistical evidence for expression change}

Bayes factors are a measure of how much the data support one hypothesis over another: the data are consistent with one gene expression probability vs. the data support there being two underlying expression probabilities, i.e. a change in gene expression has occurred. We $\log_{10}$ transform the Bayes factors. A $\log BF = 0$ means there is an equal probability for both hypotheses, whereas a $\log BF > 0$ favours Hypothesis 2 (change in gene expression) and $\log BF < 0$ favours Hypothesis 1 (no change in gene expression). Assigning each gene a $\log_{10}$ Bayes factor, enables each gene to be ranked according to the evidence supporting gene expression change given the RNA-Seq data. This opens the possibility to move away from arbitrary fold change cutoffs, that have little biological significance. 



\subsection{The more data, the stronger the statistical support can become}
\begin{figure*}[h!]
    \centering
    \begin{subfigure}[b]{0.45\textwidth}
        \centering
        \includegraphics[width=\textwidth]{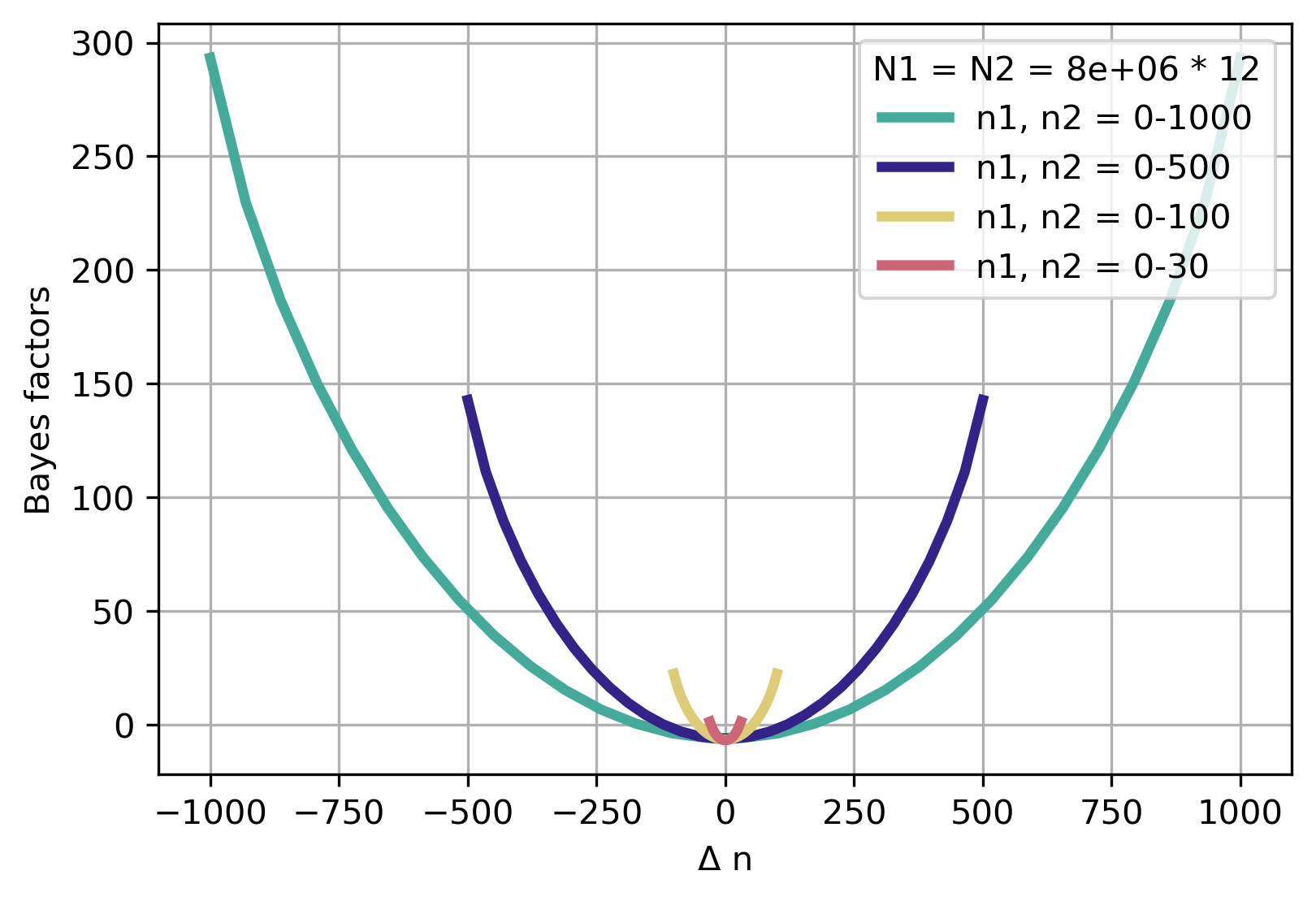}
        \centering 
        \caption{\centering}
        \label{fig:2A}
    \end{subfigure}
    \hfill
    \begin{subfigure}[b]{0.45\textwidth}   
        \centering 
        \includegraphics[width=\textwidth]{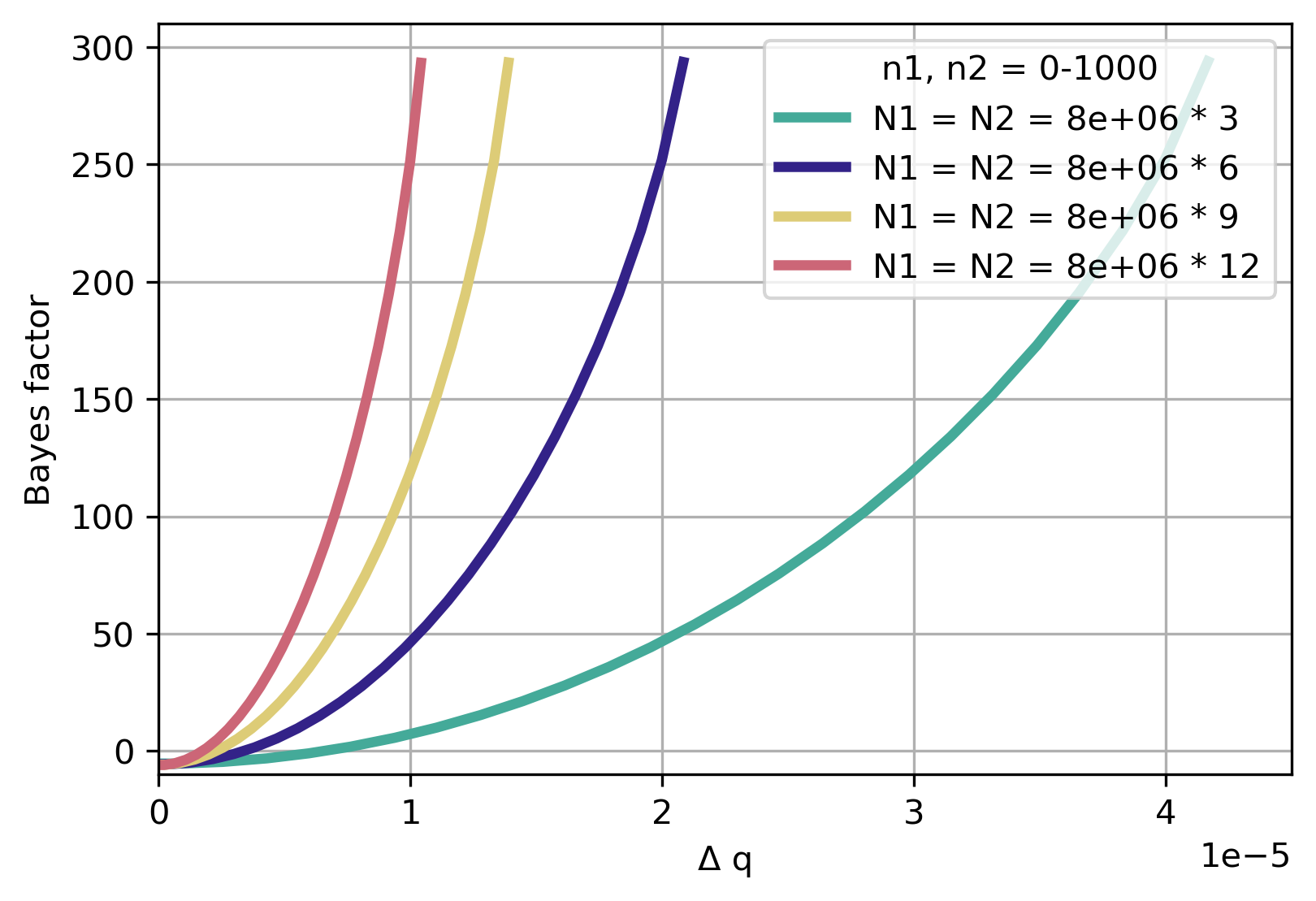}
        \caption{\centering}
        \label{fig:2C}
    \end{subfigure}

    \vspace{3mm}

    \begin{subfigure}[b]{0.45\textwidth}  
        \centering 
        \includegraphics[width=\textwidth]{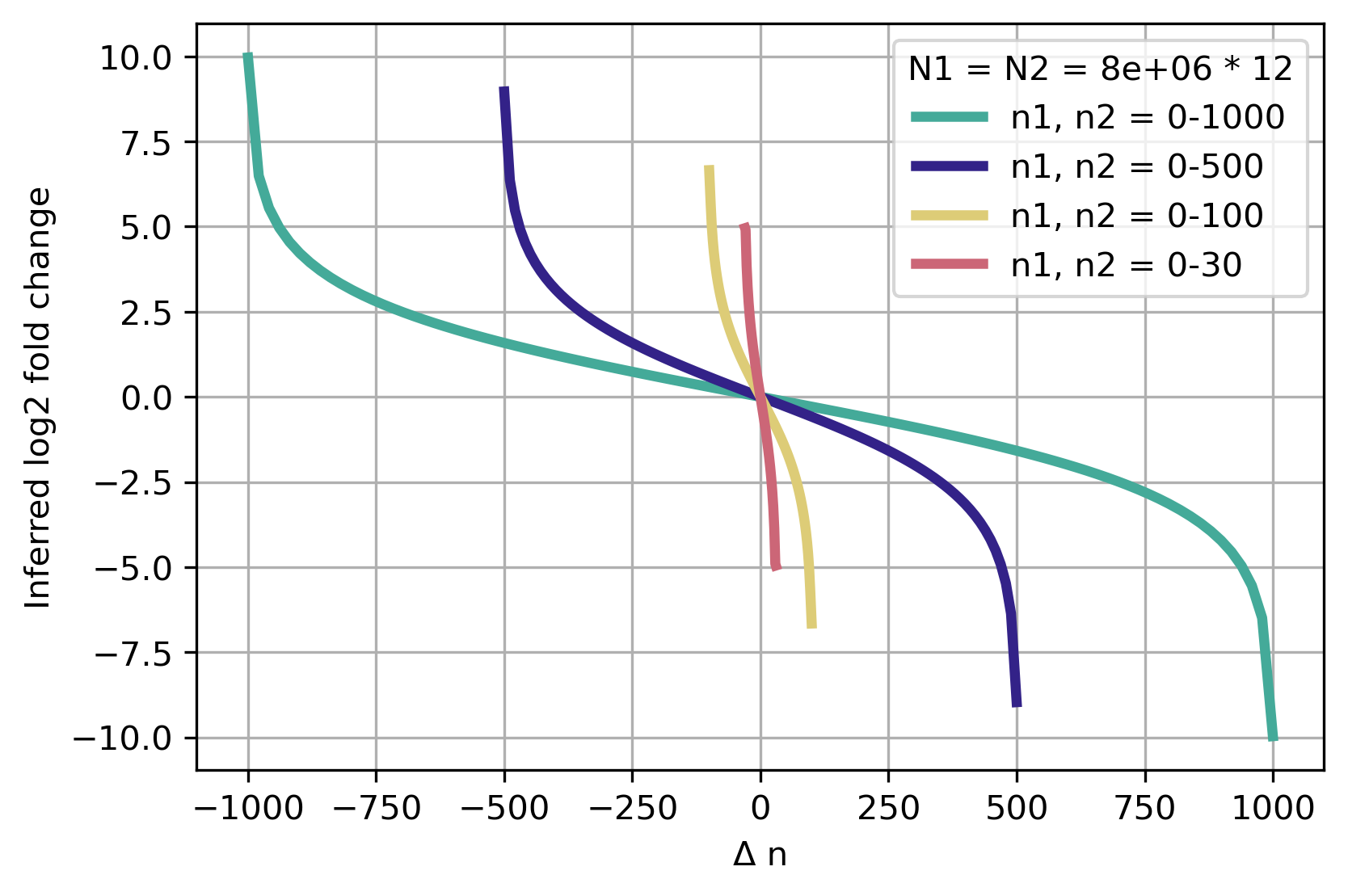}
        \caption{\centering}
        \label{fig:2B}
    \end{subfigure}
    \hfill
    \begin{subfigure}[b]{0.45\textwidth}   
        \centering 
        \includegraphics[width=\textwidth]{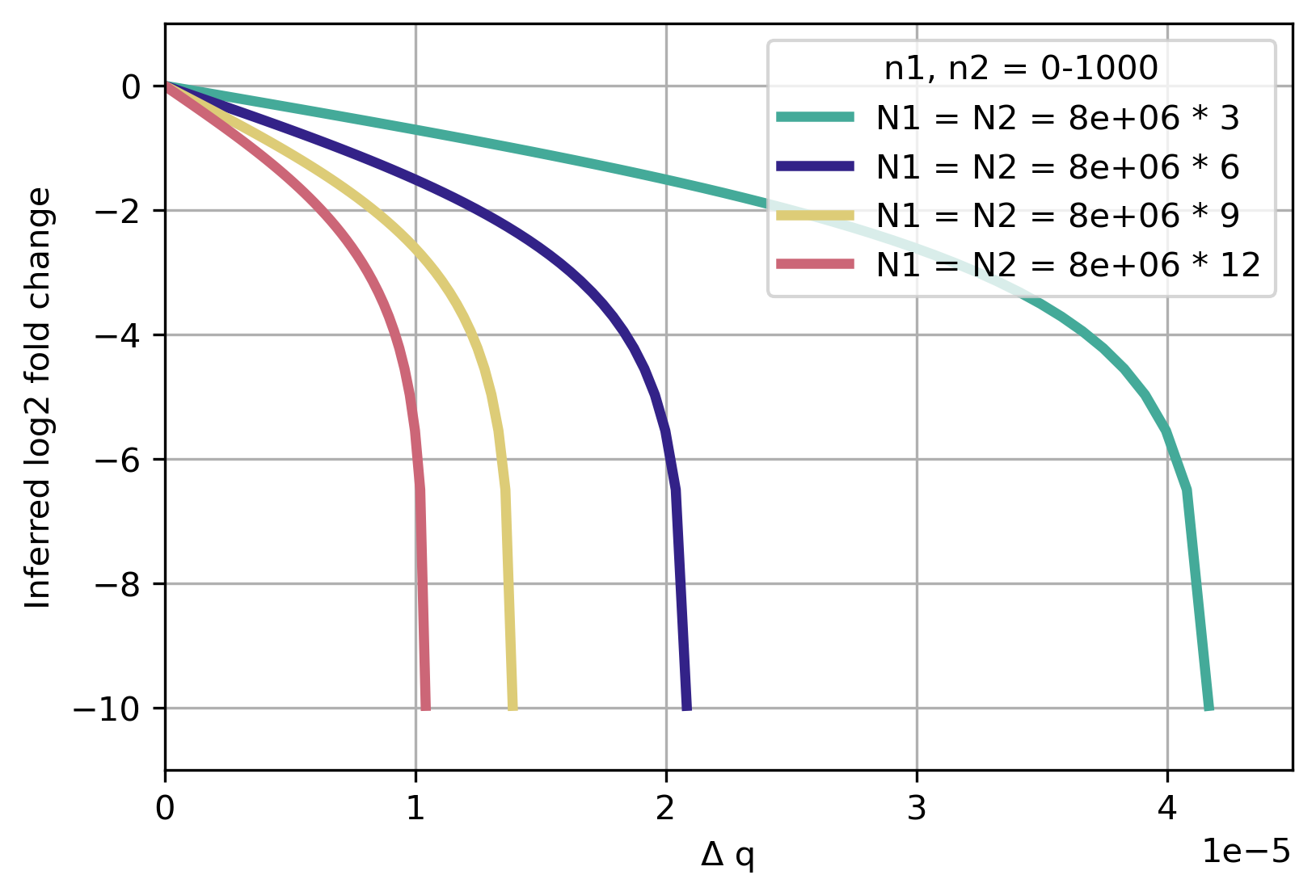}
        \caption{\centering}
        \label{fig:2D}
    \end{subfigure}
    \vspace{5mm}
    \caption[]{Bayes factors increase with the number of reads in RNA-Seq experiments. Bayes factors and inferred $\log_2$ fold change were calculated following the equations in the Results. In (a) and (c) the total number of reads in both {\it in silico} experiments is set to $8 * 10^6 * 12$. This number follows the average read depth in the RNA-Seq study of Schurch et al. \cite{Schurch_Barton_2016} and their recommended number of 12 biological replicates. In (a) and (c) the dependencies of Bayes factors and $\log_2$ fold change on the difference between the number of reads mapping to a gene in two different conditions, $n_1$ and $n_2$ ($\Delta \ n = n_2 - n_1$), are shown in different colors for different examples of $n_1$ and $n_2$. In (b) and (d) we can see the behavior of Bayes factors and inferred $\log_2$ fold change as a function of $\Delta \ q = ((n_1 + 1)/(N_1 + 2))-((n_2 + 1)/(N_2 + 2))$ for different total read depths.}
    \label{fig:2}
\end{figure*}

In Figure \ref{fig:2} we investigate the general behavior of Bayes factors and inferred $\log_2$ fold change and their relationships for different total read depths and number of reads mapping to single genes. As expected, the more data, the stronger the statistical support can become, and the more pronounced the Bayes factors. 

\section{Materials and Methods}
All presented equations in this paper can be implemented in just a few lines of code and can be used to calculate Bayes factors for differential gene expression and inferred $\log_2$ fold change values for all genes from processed RNA-Seq data (read counts). We provide a Python implementation on our \href{https://github.com/Morris-Research-Group/BDGE}{Github}, alongside with the code to reproduce the Figures of this paper.

\section{Conclusion}
Advances in sequencing technologies now allow us to monitor molecular changes within tissues and cells, offering tremendous opportunities to unravelling how gene regulation orchestrates development and responses to the environment \cite{Wang_Snyder_2009, Behjati_Tarpey_2013, Furlan_Pelizzola_2020}. The impact of RNA-Sequencing is, however, highly linked to our ability to handle the data. Reoccurring questions in this field relate to identifying and quantifying changes in RNA levels \cite{Su_SEQC/MAQC-IIIConsortium_2014, Anders_Robinson_2013, Zhou_Robinson_2014, Conesa_Mortazavi_2016, VandenBerge_Robinson_2019a, Corchete_Burguillo_2020, Koch_Winter_2018,  Costa-Silva_Lopes_2017, Costa-Silva_Lopes_2023, McDermaid_Ma_2019, Stark_Hadfield_2019,Schurch_Barton_2016, Chen_Marquez-Lago_2023, Rapaport_Betel_2013, Love_Anders_2014}. With our closed-form solution for the 2-sample problem in differential gene expression analysis, we provide an elegant statistical tool for ranking genes according to changes in RNA levels in RNA-Seq data. Despite discussions for, against or how to exactly use Bayes factors \cite{Robert_Robert_2015, Ly_Wagenmakers_2016, Kamary_Rousseau_2018} and hypothesis testing \cite{Wasserstein_Lazar_2016, Matthews_Matthews_2021, Lakens_Zwaan_2018, Amrhein_McShane_2019a} we found Bayes factors help to avoid arbitrary cutoffs \cite{Greenland_Greenland_2023} in the quantification of differential gene expression. Although other packages tackling this problem have explored the use of Bayesian statistics  \cite{Hardcastle_Kelly_2010, Leng_Kendziorski_2013, Kelter_Kelter_2021, Chen_Metzger_1998}, the exact framework and model for two-sample tests we present, is new in this analysis. We note that our results are a special case of the generalised two-sample test, previously described by Borgwardt and Ghahramani \cite{Borgwardt_Ghahramani_2009}. The closed-form solution speeds up the analysis and decreases computational expenses drastically. In future work, the framework will be tested on real RNA-Seq data and compared to existing statistical packages. Comparisons are a challenging task, as there is no ground truth given, and biological and technical fluctuations add noise to the system \cite{Schurch_Barton_2016, Chen_Marquez-Lago_2023, Rapaport_Betel_2013, Love_Anders_2014}. This is, however, where ranking genes by Bayes factors demonstrates strength, because the amount of data that is available for each gene (read counts) is taken into account. Precisely how Bayes and Laplace viewed probability \cite{Sivia_Skilling_2006}: What is the best inference we can make given the (few) data we have? Additionally, the framework may be extended for the newest technologies like single-cell RNA-Sequencing technologies and multiple-treatment comparisons. And of course, applications of this work are not limited to molecular biology, but to any two-sample test problems that can be cast into the binomial framework. \\

\subsection*{Acknowledgments}
We kindly thank Hugh Woolfenden, Pirita Paajanen, Ander Movilla-Miangolarra, and all members of the Morris Group for insightful discussions during the development of this framework and manuscript.

\subsection*{Funding}
This article is part of a project that has received funding from the European Research Council (ERC) under the European Union's Horizon 2020 research and innovation programme (Grant agreement No. 810131). G.S.S. was supported by the UK Biotechnology and Biological Sciences Research Council (BBSRC) Norwich Research Park Biosciences Doctoral Training Partnership (Grant number: BB/T008717/1).

\subsection*{Author contributions}
Conceptualization, R.J.M. and F.H.; formal analysis, R.J.M.; software, F.H., G.S.S.; writing---original draft preparation, F.H.; writing---review and editing, F.H., G.S.S., M.T., R.J.M; visualization, F.H.; supervision, R.J.M., M.T.; project administration, R.J.M.; funding acquisition, R.J.M. All authors have read and agreed to the published version of the manuscript.

\clearpage
\printbibliography

\end{document}